\documentstyle[psfig,times]{mn}
\def\H0{{\it H}$_0$}

\def\Ls{{\it L}$_\odot$}
\def\q0{{\it q}$_0$}

\def\ergps{erg~s$^{-1}$}
\def\kmpspMpc{km~s$^{-1}$~Mpc$^{-1}$}

\def\nH{$N_{\rm H}$\thinspace} 

\def\psqcm{cm$^{-2}$}
\def\ergpspsqcm{erg~cm$^{-2}$~s$^{-1}$}

\def\Zs{$Z_{\odot}$}

\def\cps{ct\thinspace s$^{-1}$}

\def\phpspsqcm{ph\thinspace s$^{-1}$\thinspace cm$^{-2}$}

\title[Chandra observation of IRAS 09104+4109] 
{Chandra detection of reflected X-ray emission from the type 2 QSO in IRAS 09104+4109} 
\author[K. Iwasawa et al] 
{\parbox[]{6.5in} {K. Iwasawa, A.C. Fabian, and S. Ettori}\\
\\
Institute of Astronomy, Madingley Road, Cambridge CB3 0HA\\ 
}
\date{}

\begin{document}

\maketitle

\begin{abstract}
We present X-ray imaging spectroscopy of the extremely luminous
infrared galaxy IRAS 09104+4109 ($z=0.442$) obtained with the Chandra
X-ray Observatory. With the arcsec resolution of Chandra, an
unresolved source at the nucleus is separated from the surrounding
cluster emission. A strong iron K line at 6.4 keV on a very hard
continuum is detected from the nuclear source, rendering IRAS
09104+4109 the most distant reflection-dominated X-ray source known.
Combined with the BeppoSAX detection of the excess hard X-ray
emission, it provides further strong support to the presence of a hidden
X-ray source of quasar luminosity in this infrared galaxy.
Also seen is a faint linear structure to the North, which coincides
with the main radio jet. An X-ray deficit in the corresponding region
suggests an interaction between the cluster medium and the jet driven
by the active nucleus.
\end{abstract}


\section{introduction}

IRAS 09104+4109 (FSC 09105+4108) is one of the extremely luminous
infrared galaxies detected in the IRAS survey and has been considered
to be a dust-enshrouded type 2 QSO (Kleinmann et al 1988; Hines \& Wills
1993; Crawford \& Vanderriest 1996; Granato, Danese \& Franceschini
1996; Soifer et al 1996; Taniguchi et al 1997; Evans et al 1998;
Hines et al 1999; Franceschini et al 2000; Tran, Cohen \& Villar-Martin 2000). 
The infrared source ($L_{\rm ir}\simeq 6\times
10^{12}h^{-2}$\Ls\footnote{\H0\ = 100 $h$ \kmpspMpc\ and \q0\ = 0.5
are assumed throughout this paper.}) has been identified with a cD
galaxy in a rich cluster at a redshift of 0.442 (Kleinmann et al
1988). Extended X-ray emission was detected with the ROSAT HRI,
and has been identified with the intracluster medium (ICM) in which 
a strong cooling flow is taking place (Fabian \& Crawford 1995; Crawford \& 
Vanderriest 1996; Allen \& Fabian 1998). 
Although the cluster emission itself is of great interest, 
it makes an X-ray investigation of the active nucleus 
difficult for X-ray telescopes with a relatively low spatial
resolution like ASCA (e.g., Fabian et al 1994) and even XMM.
The first X-ray evidence for a luminous active nucleus came from an
observation in the hard X-ray band where cluster emission is negligible.
The detection of a hard X-ray excess with BeppoSAX,
although its significance is marginal ($3\sigma$) and subject to possible 
contaminating sources in the large field of view of the PDS detector
(Frontera et al 1997), 
suggests a strongly absorbed nucleus (Franceschini et al 2000). 
The other way to access the active nucleus is to utilise 
a high spatial resolution to
separate the nucleus from the diffuse emission in an appropriate 
energy range. With the one-arcsec resolution of the X-ray telescope 
(Weisskopf, O'Dell \& van Speybroeck 1996) and
the ACIS CCD detectors (Garmire et al 2000), Chandra overcomes the difficulty.
In this letter, we present the spatially resolved X-ray source associated
with the nucleus of IRAS 09104+4109 and its spectral properties.

\section{observation and data reduction}

IRAS 09104+4109 was observed with the Chandra X-ray Observatory on 1999
November 3. The galaxy was positioned near the aimpoint of ACIS-S3,
which was operating in Faint mode. Events with the ASCA grade of 0, 2,
3, 4 and 6 were selected. The background of the ACIS-S3 chip remained 
within a factor of 2 of the mean count rate most of the time, but flared
occasionally by up to a factor of 4. We have discarded the periods of 
unusually high background events for the data of extended emission,
which left a total of 8.4 ks exposure. For a
spectrum of the nuclear point source extracted from a small
region of the detector, we consider the background variation to have no effect
and used the full exposure of 9.1 ks. 

The data were reduced using software
in CIAO 1.1.4 with the most up-to-date calibration available
from Chandra Science Center. The temperature of the focal plane
detector is estimated to be $-110^{\circ}$C, and an appropriate Fits Embedded
Function (FEF) file is selected accordingly in making the response
matrix. The energy resolution of the spectrum above 3.5 keV is about 3
per cent (or $\sim 140$ eV at 4.5 keV where a Fe K line is expected
for the galaxy's redshift) in FWHM.
X-ray spectra were extracted from the {\tt PI} column and relevant
response matrices were created by {\tt mkrmf} and {\tt mkarf}. A
spectral analysis was performed using XSPEC version 11.

\section{results}

\subsection{X-ray image}


\begin{figure}
\vspace{-4mm}
\centerline{\psfig{figure=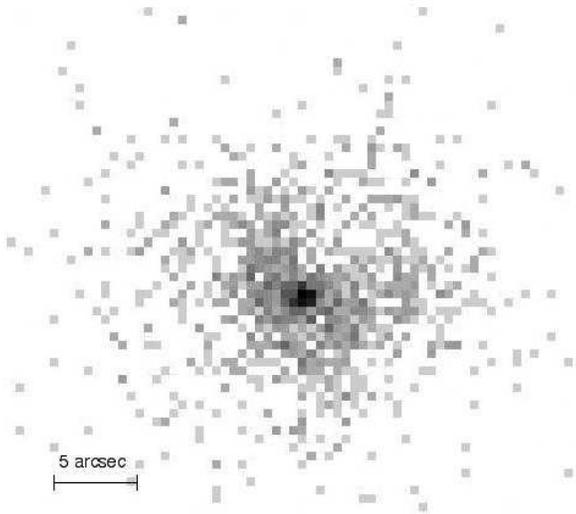,width=0.45\textwidth,angle=0}}
\caption{The 0.5--8 keV image of IRAS 09104+4109 obtained from the Chandra 
X-ray Observatory ACIS-S3. The unsmoothed full-resolution image 
is presented in logarithmic scale. In the image, North is up, East is 
to the left, and an angular scale of 5 arcsec is indicated. 
The brightest pixel of the image 
has 42 counts for a 9.1 ks exposure. Note that the North-West quarter of 
the extended emission has less X-ray surface brightness than the rest
and at a $P.A.\sim 340$ (NNW), a faint linear feature, corresponding the North 
radio lobe and hot spot (see Hines \& Wills 1993), is seen.}
\end{figure}


\begin{figure}
\vspace{-5mm}
\centerline{\psfig{figure=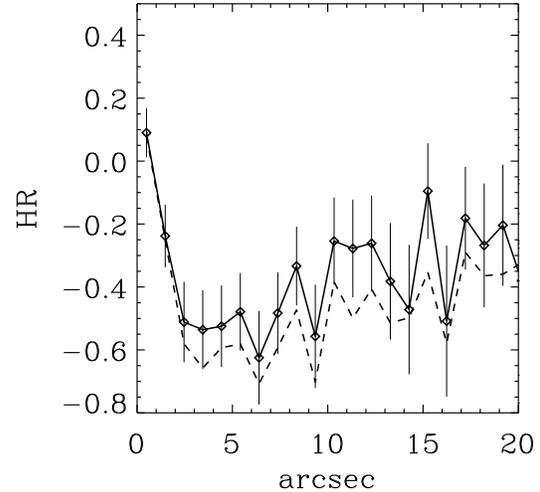,width=0.45\textwidth,angle=0}}
\caption{Radial profiles of hardness ratios, defined by 
${\rm HR=(H-S)/(H+S)}$, where H and S are integrated counts per unit area in
the hard and soft bands, respectively. The solid line represents the 
hardness ratio for H: 2--7 keV and S:0.3--0.8 keV, while the dashed line for
H: 2--7 keV and S:0.8--2 keV. The x-axis is measured from the position of the nuclear point source and the profiles have been averaged azimuthally. 
Note the hard spectrum of the nuclear source.}
\end{figure}

The full-resolution raw image (1 pixel $\sim 0.5$ arcsec) 
of the central part of IRAS 09104+4109 
in the 0.5--8 keV band is shown in Fig. 1. A bright point-like source
is seen in the middle of the X-ray image surrounded by 
extended emission of an irregular shape. The image profile of the
bright source is consistent with the point spread function. This 
point-like nature becomes clearer in the high energy band (e.g., 3--8 keV)
image, in which most of the source photons are concentrated within 2 pixels
in radius. 
The radial profile of the hardness ratios demonstrate the distinctly hard 
spectrum of the nuclear source (Fig. 2).

The position of the point source is located at $\sim 1.5$ arcsec to
the South-West of the radio core measured by Hines \& Wills (1993),
which we belive is due to an inaccurate aspect solution in the Chandra data. 
Notable is a faint linear feature at $\sim 10$ arcsec from the 
point source extending to the NNW (at a position angle of
$\sim 340^{\circ}$). The feature points to the bright central source
and, assuming the central source to coincide with the active nucleus, 
it matches the radio structure of the preceding North radio lobe and 
hot spot, found in the VLA images (Kleinmann et al 1988; Hines \& Wills 1993).
However, this X-ray feature is very faint and it is difficult to assess
its significance due to the non-uniform background of the complex cluster 
emission (about $2\sigma $ excess compared to the surroundings). 
Our estimate of the flux at 1 keV from the jet is
$(3\pm 2)\times 10^{-15}$\ergpspsqcm, more than one order of magnitude above
the radio power measured with VLA (Hines \& Wills 1993). 
The radio jet is believed to emanate close to the plane of the sky and 
its bulk velocity is $\sim 0.1c$ (Hines et al 1999). 
No detection of an optical jet is reported.
This X-ray feature could be Compton-upscattered photons of 
beamed radiation from the nucleus in the mildly relativistic electrons
(Celotti, Ghisellini \& Chiaberge 2000).

Evidence for an interaction between the AGN jet and the ICM is 
found in the X-ray image. It is interesting to note that the radio 
jet is rather slim compared to the large opening angle 
($\sim 90^{\circ}$) of the low X-ray surface-brightness region, 
unlike the other examples such as Hydra A cluster and Perseus cluster.
A detailed investigation of the cluster emission will appear in the
forthcoming paper.

It should also be noted that the NW quarter of the inner part of the 
extended emission is significantly lower in X-ray surface brightness
by 30--40 per cent than the rest. This X-ray deficit was probably viewed 
as a ``hole'' in the ROSAT HRI image (Fabian \& Crawford 1995;
Crawford \& Vanderriest 1997). Since this region lies along the direction of
the radio/X-ray jet, it is plausible that the ICM has been 
affected by the jet and thus the low surface-brightness region is
similar to the X-ray cavities 
seen also by Chandra in other clusters harbouring 
a strong radio source in their
centre (e.g., Hydra A cluster, McNamara et al 2000; 
Perseus cluster, Fabian et al 2000a). However, it is interesting to note 
the relatively wide opening angle ($\sim 90^{\circ}$) of the cavity
despite the rather slim radio jet in IRAS 09104+4109. 
The optical emission plume
(Kleinmann et al 1988; Hutchings \& Neff 1988; Soifer et al 1996; Crawford \& Vanderriest 1997;
Armus et al 1999; Tran et al 2000) is located at the edge of this
X-ray cavity and a strongly blueshifted component of emission-line 
gas at the nucleus, supporting the idea of a central outflow.

Here, we mention the cluster emission briefly, and the further details
will appear elsewhere. The large-scale X-ray envelope is elongated along
the East-West direction, which is consistent with the flattened galaxy
distribution of the cluster (Kleinmann et al 1988), 
while the X-ray morphology of the inner part is more complex.
A significant radial temperature gradient is found: the temperature of
$7.8\pm 1.5$ keV found at radii of 200 kpc drops to $3.3\pm 0.3$ keV
near the nucleus, when fitted to a single-temperature thermal emission model. 
This clearly indicates that strong cooling of the ICM is
taking place in the dense core of the cluster, as previously suggested
from the X-ray imaging and spectral analysis 
(Fabian \& Crawford 1995; Allen \& Fabian 1998; Ettori \& Fabian 1999).

\subsection{X-ray spectrum of central source}


\begin{figure}
\centerline{\psfig{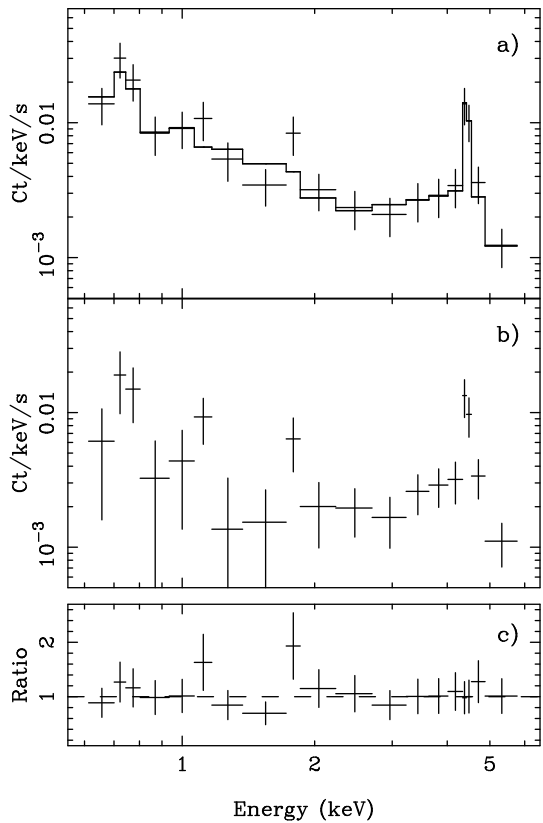}}
\caption{The Chandra ACIS-S spectrum of the central source of IRAS
09104+4109: a) the data with no background subtraction, 
and fitted with a model consisting of the thermal emission model,
cold reflection contniuum ({\tt pexrav}) and a gaussian; b) the 
same data as a) but subtracted by the data taken from the 
$1\leq r\leq 2$ arcsec annulus normalized by geometrical area;
c) the residual of the fit given in a) represented by ratio between the data and the model. a  A strong line
feature seen at 4.5 keV is a redshifted Fe K$\alpha $ line (6.4 keV),
which has an equivalent width of $\sim 1$ keV with respect to the 
hard continuum.}
\end{figure}

The spectral data of the central source were collected from a circular
region with a 1-arcsec radius. Over 80 per cent of photons 
from a point source are expected to fall in this region.
The count rate is 0.022 \cps in the 0.6--7 keV band and the spectrum is 
shown in Fig. 3. 
A prominent feature in the spectrum is the strong line at 4.5 keV on the
hard excess emission, which is identified with Fe K$\alpha $ at
a rest energy of 6.4 keV. 

The hard component above 3 keV can be described by a  
power-law absorbed by \nH $\sim 3\times 10^{23}$\psqcm\ when a reasonable
photon index is assumed (e.g., $\Gamma = 2$). However, extrapolating this
model to higher energies fails to explain the BeppoSAX detected flux at
20--30 keV by a factor of $\sim 30$. We therefore rule out the possibility
of an absorbed continuum and attribute this component
to a reflection continuum from cold matter. This interpretation is more
favourable also on account of the strong Fe K line. The XSPEC model,
{\tt pexrav} (Magdziarz \& Zdziarski 1995) appropriate
for a reflection spectrum from
neutral matter, provides a good fit to the data apart from the Fe K line.
For the illuminating source, a power-law spectrum with $\Gamma = 2$ and
no high-energy cut-off is assumed. 
The quality of the fit is insensitive to the choice
of the power-law slope in a reasonable range.
The Fe K line is found at the rest-energy of $6.40\pm 0.07$ keV.
The line width is $60^{+150}_{-60}$ eV in gaussian dispersion 
and the line intensity is $7.2^{+6.1}_{-3.3}\times 10^{-6}$\phpspsqcm, 
corresponding to a redshift-corrected equivalent width 
of $1.1^{+0.9}_{-0.5}$ keV. Note that the Fe K line of the cluster gas
is at a clearly higher energy at $\sim 6.7$ keV.

Below 2 keV, a steep rise is seen in the spectrum towards lower energies.
The spectrum in this energy range is sensitive to the background subtraction. 
The cluster emission increases its surface brightness steeply towards
the centre, which makes an accurate estimate of the contribution from the 
cluster to the nuclear spectrum difficult.
If no background is subtracted, the soft X-ray emission can be described
by the thermal emission model, {\tt MEKAL}, with a temperature of 
$kT=1.7^{+0.8}_{-0.3}$ keV and Solar metallicity (Fig. 3a). 
Higher metallicity (2--3 \Zs) is favoured by the data when left as a 
free parameter, although the constraint is rather loose.
No excess absorption above the Galactic value,
(\nH $\simeq 1.8\times 10^{20}$\psqcm, Dickey \& Lockman 1990;
Murphy et al 1996), is required.
The temperature is significantly lower than the cluster emission
near the nucleus ($kT\approx 3.3$ keV).
This suggests that a softer spectrum of the the nuclear source composes
a significant fraction in the energy range.
This is supported by the fact that the soft X-ray image shows a narrow
peak at the same position as the hard X-ray peak.
The excess of the nuclear emission, in fact, 
appears to be largely due to strong emission
lines, particularly a strong feature at 0.75 keV
($\sim 1.1$ keV in the rest frame and due to 
the Fe L complex and ionized Ne).
If the background data are taken from an annulus of 
the immediate surrounding ($1\leq r\leq 2$ arcsec in radii), 
subtracting this background makes no effect on the spectrum above 3 keV,
but reduces a count rate in the 0.6--2 keV band by 44 per cent and
leaves possible emission features (Fig 3b).
This soft X-ray emission could be the emission from photoionized gas
at the inner nucleus.
However, we tentatively fit the data with no background 
to the thermal emission model here and 
will discuss the origin in the Discussion section.

A spectral fit with the components described above is shown in Fig. 3a
where the resulted $\chi^2$ value is 8.5 for 14
degrees of freedom (Fig. 3c). Placing a reasonable amount of absorption
(a few times of $10^{21}$\psqcm\ in \nH, e.g., equivalent absorption
to the NLR reddening) on the reflection component does not make any
significant difference in the quality of the overall fit or spectral
parameters of the thermal component.  The observed 0.5--7 keV flux is
estimated to be $2.5\times 10^{-13}$\ergpspsqcm, which can be
translated to the rest-frame 0.7--10 keV luminosity of $6\times
10^{43}h^{-2}$\ergps.

\section{Comparison with the BeppoSAX data}
The BeppoSAX PDS data (14--40 keV range) were taken from the archive
to compare with the Chandra data.
Franceschini et al (2000) attributed the hard X-ray emission detected
with the BeppoSAX PDS to a strongly absorbed, transmitted component.
Using the spectral fit to the Chandra data, we find that this conclusion
depends on the assumed spectral slope of the source illuminating the 
reflecting matter. Given the low signal-to-noise ratio of the PDS data,
no good constraints can be obtained for the spectral slope however.
We thus show only two cases of $\Gamma = 2$ and $\Gamma = 1.4$ for 
a primary source spectrum.

When $\Gamma = 2$, typical for a quasar, is assumed,
even with an almost face-on setting of the reflection slab in 
{\tt pexrav} model (which yields the hardest spectrum above 10 keV),
the PDS data lie a factor of $\sim 5$ above the extrapolation of the 
cold reflection model for the Chandra nuclear spectrum.
This suggests the presence of an extra high energy component due
to transmitted radiation, as proposed by Franceschini et al (2000).
If an absorbed power-law is fitted to this excess component, the
best-fit value of the column density is found to be
$3.3\times 10^{24}$\psqcm\ (cf., Franceschini et al 2000 obtained 
$6.6\times 10^{24}$\psqcm). We note a large statistical error to
the \nH value and further ambiguities arising from Compton scattering
and iron metallicity in such a high column density range 
(Matt, Pompilio \& La Franca 1999; Wilman \& Fabian 1999). 
Despite all the uncertainties, 
the Thomson depth of the X-ray absorber should exceed unity.
The apparent lack of the transmitted component in the Chandra energy
range sets the lower limit of the column density to be 
$2\times 10^{24}$\psqcm, while the detection of the hard X-ray excess
means \nH\ is no larger than $10^{25}$\psqcm. A joint fit to the Chandra
ACIS and BeppoSAX PDS spectra with this model gives 
$\chi^2=10.0$ for 18 degrees of freedom (see Fig. 4).

In the case of $\Gamma = 1.4$, which is on the flatter side of the 
photon index distribution of quasars (e.g., Reeves \& Turner 2000),
the PDS data are explained well by cold reflection best-fitting
the ACIS data (as mentioned in the previous section, the quality of the
fit to the ACIS data is not sensitive to the selection of 
slope) with $\chi^2=10.8$ for 20 degrees of freedom.
In this case, no transmitted component is required and the column density
of the X-ray absorber exceeds $10^{25}$\psqcm.


\begin{figure}
\centerline{\psfig{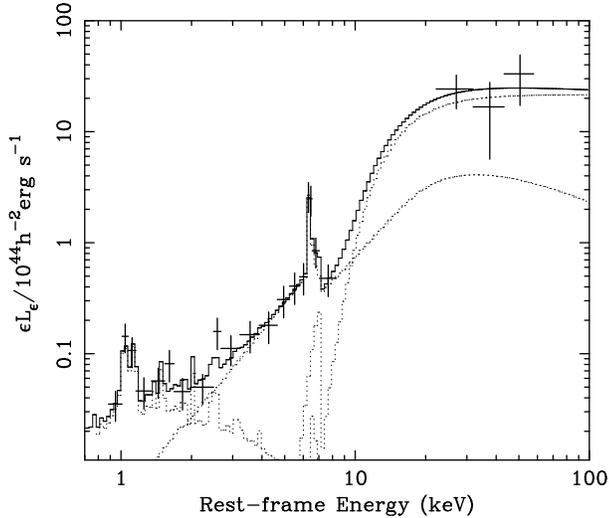}}
\caption{The Chandra data of the nuclear source of IRAS
09104+4109 and the BeppoSAX PDS data with the best-fit model (solid 
histogram) consisting of thermal emission, cold reflection and highly absorbed 
emission components (dotted line histograms). The values of the data and model
have been converted to those of luminosity assuming redshift
of 0.442, \H0 = 100 \kmpspMpc and \q0=0.5. The energy scale has been
corrected for the galaxy's redshift.}
\end{figure}

\section{discussion}

\subsection{Buried QSO}

A distinctive cold reflection feature is observed from the nucleus
of IRAS 09104+4109 in the Chandra spectrum
whilst no primary radiation is visible.
Detection of the primary emission at higher energies with BeppoSAX
(Franceschini et al 2000)
depends, as demonstrated above, on the assumed spectrum of the source.

If the PDS-detected X-rays are entirely due to reflection, only a lower
limit of the primary source luminosity can be obtained which is 
a minimum requirement to produce the observed luminosity 
through cold reflection.
The albedo ($\eta $) in the 2--10 kev band 
is calculated using the {\tt pexrav} model for two 
inclination angles, $i=20^{\circ}$ and 60$^{\circ}$, for the reflecting 
slab subtending $2\pi$ in solid angle.
For $\Gamma = 1.4$, $\eta = 0.066$ ($i=20^{\circ}$) and 0.052
($i=60^{\circ}$). Therefore the luminosity of the primary source is 
larger than $5\times 10^{43}\eta^{-1}\simeq 7.6\times 10^{44}$\ergps\ and
$9.6\times 10^{44}$\ergps\ for respective inclinations.
Since the fraction of reflecting surface visible to us is likely 
to be less than unity, 
the true luminosity of the primary source should be larger than these
values.

We discuss below the case in which the primary source has a power-law
of $\Gamma = 2$ and hence its transmitted radiation is detected with the PDS.
The 2--10 keV luminosity of the primary source corrected for the absorption
(and effects of Compton scattering, Matt et al 1999) 
is estimated to be $7.4\times 10^{45}\gamma h^{-2}$\ergps 
($\gamma\sim 1$ for a spherical obscuration, and it can go up to
$\sim 3$ if absorption occurs at a small cloud in the line of sight
when \nH $=3\times 10^{24}$\psqcm).
The observed-to-intrinsic luminosity ratio in the 2--10 keV band
is then $0.67\gamma^{-1}$ per cent for IRAS 09104+4109.
If the ratio of the 2--10 keV to bolometric luminosities
is assumed to be $f_{\rm HX/Bol}=0.05f_{\rm HX/Bol,0.05}$, typical
for quasars (e.g., Elvis et al 1994),
the bolometric luminosity is $1.5\times 10^{46}f_{\rm HX/Bol,0.05}^{-1}
\gamma h^{-2}$\ergps.
This is comparable to $L_{\rm ir}=2.3\times 10^{46}h^{-2}$\ergps, and 
consistent with a picture where the bulk of the infrared emission is
powered by a buried quasar through dust reradiation.
The radio source has an intermediate power between FRI and FRII
(Kleinmann et al 1988; Hines \& Wills 1993). However, without the
nuclear obscuration, the wide
band energy distribution of IRAS 09104+4109 would be closer
to radio quiet AGNs.

The cold gas reflecting the observed X-rays 
is probably part of the clouds occulting the central source. 
An X-ray absorber with as large a column density as found in 
IRAS 09104+4109 is likely to be compact, probably parsec-scale
in radius as for nearby Compton-thick Seyfert 2 galaxies (Matt 2000). 
The flat energy distribution in the infrared band indicates that 
hot dust at near sublimation temperature ($\sim 1000$ K) is
present and the inner radius of the dusty torus is 
the order of 1 pc (Granato et al 1996; Taniguchi et al 1997).
The mid-infrared spectrum obtained from ISOCAM CVF {(Taniguchi et al 1997)
fits well with 
the compact dusty torus model of Pier \& Krolik (1992) with an edge-on
view (also see Granato \& Danese 1994; Granato et al 1996) 
whereas optical polarization studies
imply a moderately inclined torus (Hines et al 1999; Tran et al 2000).
The characteristic dust temperature is about 120 K (e.g., Kleinmann et al
1988).
No IRAS detection at 100$\mu $m and the new SCUBA limit at 850$\mu $m 
(Deane \& Trentham 2000) rule out the presence of cool dust, 
which is a common feature of local ULIGs.
No detection of CO (Evans et al 1998) is consistent with a relatively
small obscuring torus with a modest mass.
A large covering factor ($\geq 0.9$) is favoured by all the observations.

The soft X-ray emission seen in the nuclear spectrum 
is probably due to photoionized gas in the inner nucleus. The highly polarized
biconical reflection nebula imaged by HST (Hines et al 1999) 
is however not a likely source, since scattering by dust rather than electrons
appears to be a 
dominant mechanism to induce the high polarization (Tran et al 2000).
The inner wall of the obscuring torus is exposed to the 
intense radiation from the primary source and thus expected
to be highly ionized. Provided the optical depth of the ionized gas is 
small, the Fe-L emission bump can be very strong (e.g., Band et al 1990).
A sub-parsec scale ionized disk has indeed been imaged 
with the VLBA in the nucleus of the 
nearby Compton-thick Seyfert 2 galaxy NGC1068 (Gallimore, Baum \& O'Dea
1997), and they predicted strong soft X-ray emission lines based on
a photoionization computation. 
In this case, the luminosity of the emission from the photoionized gas
should be larger than observed ($\sim 3\times 10^{42}h^{-2}$\ergps\ in 
the 0.5--2 keV band),
as significant absorption is likely
to occur during the escape from the nuclear region.

A possible alternative is hot stars near the nucleus, 
inferred from a Wolf-Rayet 
feature seen in the optical spectrum of the nuclear region (Tran et al 2000). 
However, the luminosity of this component alone is one
order of magnitude more luminous than the nearby Wolf-Rayet galaxies
observed with ROSAT (Stevens \& Strickland 1998), and the high
metallicity implied from the strong emission-line features is also unusual
(but see Buote 2000).

\subsection{Implications for other luminous IR galaxies}

IRAS 09104+4109 is so far the only hyper-luminous infrared galaxy detected
in X-rays (apart from PG1634+706 which is an unobscured quasar, see Nandra
et al 1995 for the ASCA result),
and the most distant reflection-dominated spectrum X-ray source known
with a clear detection of the Fe K line. 
The small X-ray reflection fraction ($\sim 0.7$ per cent in the 2--10 keV 
band; $\sim 0.1$ per cent at the rest energy of 2 keV) 
estimated for IRAS 09104+4109 is still consistent with the
limit obtained from the previous observations for other luminous
infrared galaxies at high redshift (see Fabian et al 1997; Ogasaka et al
1997 for F15307+3252; Wilman et al 1998). Chandra and XMM are capable of
detecting reflected X-ray light from F15307+3252, if the reflection fraction
is similar to IRAS 09104+4109. Although a $2\sigma $ detection 
of the lensed object IRAS F10214+4724 with ROSAT HRI was reported 
(Lawrence et al 1994), 
an 80 ks ASCA observation failed to detect it (Iwasawa 2000). The prospect
of X-ray detection of this object depends on the lensing geometry.
It should be noted that IRAS 09104+4109 is peculiar among the ULIG population
being in a cluster environment, and how common this type of objects is is 
unclear (e.g., Deane \& Trentham 2000).

Chandra observations of the SCUBA sources have revealed 
no correlation between the sub-millimetre and X-ray source populations
(Fabian et al 2000b; Hornschemeier et al 2000) with two exceptions 
(Bautz et al 2000). If ULIGs at high redshift are powered predominantly
by AGN, as suspected by Trentham, Blain \& Goldader (1999); Trentham (2000), 
their nucleus is Compton-thick and the X-ray reflection fraction 
is indeed as low as that in IRAS 09104+4109. On the other hand, 
the sources with a small X-ray optical depth, that are preferably detected 
in X-ray, may tend to have hotter dust (Wilman, Fabian \& Gandhi 2000)
which deters submillimetre detection.

\section*{Acknowledgements}

We thank all the members of the Chandra team for building and
operating the satellite and developing the data analysis
software. Neil Trentham, Steve Allen and Carolin Crawford are thanked
for helpful discussion. The Royal Society (ACF,SE) and PPARC (KI) are
thanked for support.

\end{document}